\documentclass[journal]{IEEEtran}
\usepackage{ifpdf}

\usepackage{amsmath}
\usepackage{algorithm}
\usepackage[noend]{algpseudocode}

\makeatletter
\def\BState{\State\hskip-\ALG@thistlm}
\makeatother

\usepackage{cite}

\ifCLASSINFOpdf
  \usepackage[pdftex]{graphicx}
   \DeclareGraphicsExtensions{.pdf,.jpeg,.png}
\else
\fi
\usepackage{lipsum}

\usepackage{amsmath}
\usepackage{amssymb}
\usepackage{flexisym}
\usepackage{array}
\newcolumntype{C}[1]{>{\centering\let\newline\\\arraybackslash\hspace{0pt}}m{#1}}

\usepackage{color}
\definecolor{dblue}{rgb}{0,0,0.8}

\usepackage{xcolor}

\usepackage{multirow}

\usepackage{subfig}
\usepackage{graphicx}
\usepackage{fixltx2e}

\usepackage{stfloats}
\usepackage{nomencl}

\usepackage{setspace}
\makenomenclature

\begin{document}
%



\title{Techno-Economic Assessment of Net-Zero Energy Buildings: Financial Projections and Incentives for Achieving Energy Decarbonization Goals}


%

\author{Hamed Haggi,~\IEEEmembership{Member,~IEEE}, James M. Fenton
\thanks{The authors are with the University of Central Florida's Florida Solar Energy Center, FL, USA. Emails: hamed@ucf.edu, jfenton@fsec.ucf.edu}
}

\maketitle

\begin{abstract}

Recent advancements, net-zero emission policies, along with declining costs of renewable energy, battery storage, and electric vehicles (EVs), are accelerating the transition toward cleaner, more resilient energy systems. This paper conducts a comprehensive techno-economic analysis of Net-Zero Energy Buildings (NZEBs) within Florida's energy transition by 2050. The analysis focuses on the financial advantages of integrating rooftop photovoltaic (PV) systems, battery storage, and EVs collectively compared to reliance on grid electricity for both existing and newly built homes in Orlando, Florida. By leveraging federal incentives like the Investment Tax Credit and considering energy efficiency improvements, residents can achieve significant savings.  Simulation results show that existing homes with a 9.5 kW PV system and 42.2 kWh battery are projected to generate positive returns by 2029, while newly constructed homes meet this threshold as early as 2024. Also, rooftop solar used to charge an EV can save up to \$100 per month for residents compared to gasoline. Combining PV and battery storage not only lowers electric bills but also enhances grid independence and resilience against grid outages. Beyond individual savings, NZEBs contribute to grid stability by reducing electricity demand and supporting utility-scale renewable applications. These advancements lower infrastructure costs, help Florida residents and utilities align with national decarbonization goals, retain approximately \$23 Billion within the state and foster progress toward a sustainable, low-carbon future.

\end{abstract}

\vspace{2mm}
\begin{IEEEkeywords}
Battery Energy Storage, Energy Decarbonization, Electric Vehicles, Energy Savings, Florida Energy Transition, Net-Zero Energy Buildings, Photovoltaic Systems, Techno-Economic Analysis.
\end{IEEEkeywords}

\IEEEpeerreviewmaketitle

\section{Introduction}
\IEEEPARstart{T}{he} global energy system contributes approximately to 75\% of greenhouse gas emissions, positioning it as a critical factor in addressing climate change. Achieving Net Zero Emissions by 2050 requires a profound transformation of energy production, transportation, and consumption across the United States \cite{anika2022prospects}. This transformation must include a rapid acceleration of clean energy technologies during the decade leading to 2030. While renewable energy deployment has grown substantially in recent decades, significant barriers remain to achieving widespread grid integration, particularly in Florida’s energy market.

Despite the implementation of more energy efficiency codes, Florida’s building sector remains a major source of emissions. Addressing the energy challenges within the state requires overcoming several critical obstacles: ensuring the resilience and security of solar energy as it integrates into the grid, demonstrating the economic value of energy storage, and reliably interconnecting solar resources. Additionally, optimizing solar system performance, leveraging transportation electrification for grid support, improving building energy efficiency, and fostering technological innovation and workforce development are all essential to achieving a decarbonized and resilient energy future for Florida.

\begin{figure}[b]
\centering
\footnotesize
\captionsetup{singlelinecheck=false,font={footnotesize}}
	\includegraphics[width=3.45in]{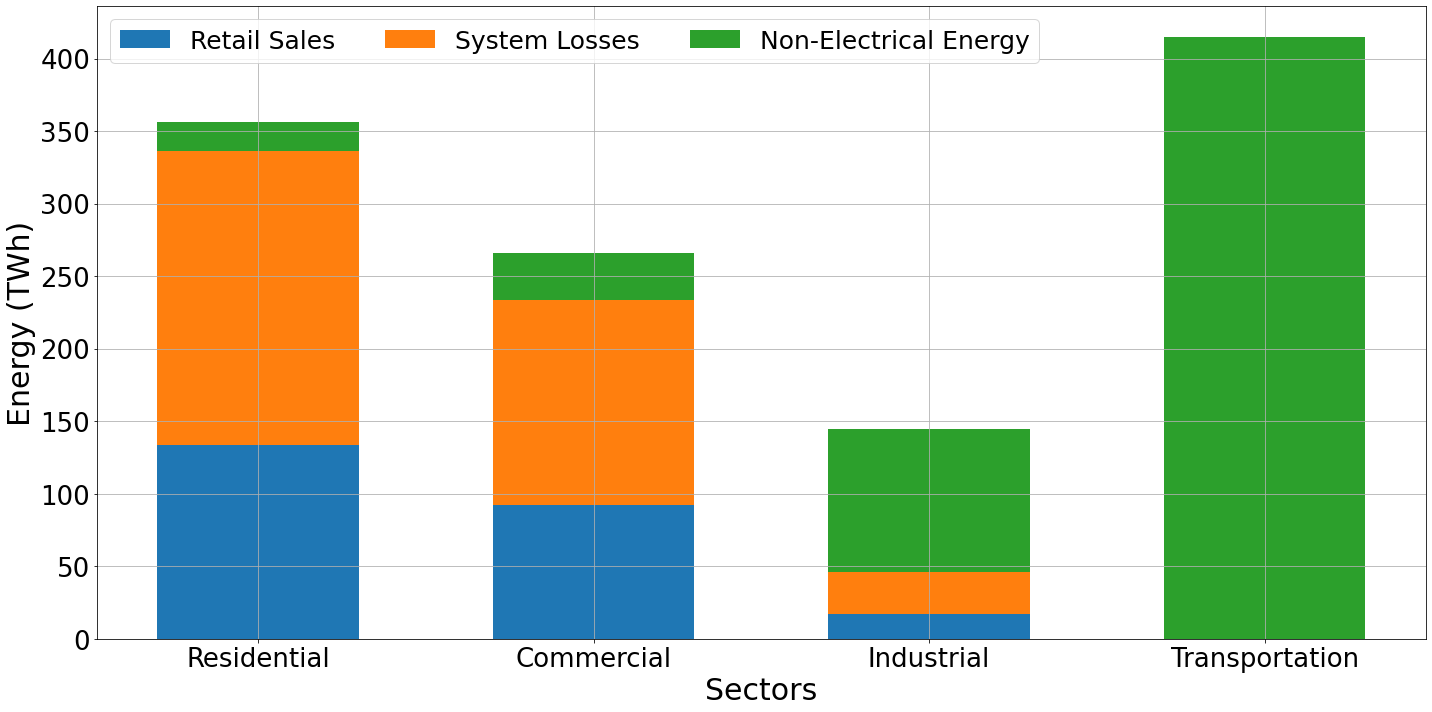}
	\caption{Florida's net energy consumption (Total 1182 TWh) considering retail sales, electrical and non-electrical system losses.}
    \label{Piechart_Fig1}
\end{figure}

In 2020, Florida’s net end-use energy consumption across the Residential, Commercial, Industrial, and Transportation sectors (presented in Figure \ref{Piechart_Fig1}) totaled 809 TWh, with 242 TWh attributed to electricity retail sales. Accordingly, energy losses in these sectors' electrical systems reached 373 TWh (resulting in a combined end-use energy consumption of 1,182 TWh)  reducing the overall efficiency of electricity delivery to just 40\%. The residential sector experienced significant inefficiencies, with electric system losses (203.1 TWh) surpassing actual electricity consumption (153.4 TWh). Transitioning to on-site solar generation and battery storage could eliminate these system losses while preserving residential electricity consumption, providing backup power, and contributing to Florida's clean energy transition. Such advancements would not only improve system efficiency but also align completely with decarbonization targets.

Reducing electricity demand in residential buildings is crucial and can be achieved through various strategies, including energy-conserving architectural designs, adoption of high-performance appliances, optimization of heating, ventilation, and air conditioning (HVAC) systems, promotion of energy-aware resident behavior, and integration of intelligent control mechanisms \cite{wu2018selecting}. Energy-efficient architectural design—whether for new constructions or retrofits—enhances building envelopes by increasing thermal insulation, incorporating advanced window glazing, and employing reflective or eco-friendly roofing technologies, all aimed at mitigating solar heat gain \cite{li2019energetic}.

Research on Net Zero Energy Buildings (NZEBs) equipped with photovoltaic (PV) and battery systems has often focused on single-day operations and assessed feasibility using financial metrics. Some studies also consider hybrid systems that include EVs to improve resilience during prolonged outages \cite{gorman2023county}. For example, Kim et al. \cite{kim2023economic} explored the economic feasibility of implementing NZEBs in U.S. homes using heat pump systems alongside federally supported solar and geothermal technologies. They highlighted the challenges of achieving "net-zero emission" targets and analyzed the payback periods of various NZEB scenarios, considering future technological and policy changes needed to meet the 2050 goals. Sohani et al. \cite{sohani2023techno} examined the benefits of PV and building-integrated PV thermal hybrid systems for electricity generation, utilizing excess electricity to drive hot and cold water storage systems in residential buildings in Tehran. Kleinebrahm et al. \cite{KLEINEBRAHM2023} conducted a techno-economic analysis to identify the most cost-effective PV solutions for achieving self-sufficient energy supply in buildings across Central Europe. Their simulations indicated that 53\% of the 41 million single-family residences have the potential to operate independently from the power grid, with projections suggesting this figure could rise to 75\% by 2050. Cucchiella et al. \cite{cucchiella2018solar} assessed the economic viability of Italian homes equipped with 3-kW PV and lead-acid battery storage systems. Zakeri et al. \cite{zakeri2021policy} explored policy incentives for energy efficiency improvements in U.K. residential buildings, aiming to reduce electricity consumption by 84\% using an optimized PV and battery storage system. Hale et al. \cite{hale2018integrating} focused on the future challenges of Florida's power system with high PV penetration and battery storage technologies, considering total residential, commercial, and utility loads, along with flexible options like battery storage and demand response to enhance PV integration and reduce carbon emissions. Aniello et al. \cite{aniello2021micro} performed a techno-economic assessment for German households by considering PV and battery storage systems, utilizing financial metrics such as NPV, cash flows, simple payback, and internal rate of return. Alipour et al. \cite{alipour2022exploring} investigated the motivations and concerns of residential customers in Queensland, Australia, regarding PV and battery adoption, finding that upfront costs and reliability concerns were key factors. Tervo et al. \cite{tervo2018economic} conducted an economic analysis of residential PV systems coupled with lithium-ion battery storage across 50 U.S. states, discovering that optimally sized PV-battery systems with net metering were more cost-effective than stand-alone PV systems. Mohamed et al. \cite{mohamed2021comprehensive} presented a comprehensive techno-economic tool for U.K. residential PV-battery systems, considering real-time battery control, capacity degradation, and investment profitability using metrics like NPV and return on investment. Koskela et al. \cite{koskela2019using} evaluated the profitability of optimally sized PV and battery storage systems for apartment buildings and detached houses in Finland, including economic analyses related to equipment sizing in response to tariff and incentive changes in power distribution systems. William et al. \cite{william2022enviro} analyzed the potential of energy-efficient building solutions in various hot and humid climates through parametric analysis, finding that installed PV reduced energy consumption and mitigated indoor thermal discomfort. Way et al. \cite{way2022empirically} generated empirically validated cost forecasts for technologies impacting the energy transition period, exploring different scenarios and finding that previous projections overestimated renewable technology costs. Kunwar et al. \cite{kunwar2023performance} investigated the efficacy of active insulation systems to enhance residential building efficiency, aiming to mitigate energy loss and reduce peak demand. Further information on NZEBs, strategies for retrofitting existing buildings, increasing renewable penetration at the customer level, and relevant policies can be found in \cite{ohene2022review}, \cite{zhang2022grid}, and \cite{ahmed2022assessment}.

Despite these studies, there is a lack of comprehensive analysis on the economic and technical benefits of NZEBs over the next 30 years that encompass PV systems, energy efficiency enhancements, stationary battery storage systems, and vehicle-to-home applications collectively. Such an analysis should consider important aspects from both consumer and utility perspectives like monthly savings for residents, the point at which residential renewable electricity becomes more cost-effective than grid electricity, and the incentives for peer-to-peer energy exchanges among residents, all while ensuring reliable energy supply even during emergencies. Additionally, it should explore the advantages NZEBs offer to utilities. To this end, the main contributions of this paper are: 

\begin{itemize}
    \item A comprehensive 30-year Techno-Economic evaluation of NZEBs was conducted in Florida (for existing and newly built homes), uniquely combining PV systems, energy efficiency improvements, battery storage, and vehicle-to-home (V2H) applications. This holistic study fills a significant research gap by considering these technologies collectively, offering new insights and incentives for both consumers and utilities' decision-making. 
    \item An empirical evaluation was conducted using real-world energy consumption data from both existing and retrofitted homes in Florida. This empirical approach yields practical insights into NZEB performance in Florida's specific climate and energy context, providing valuable guidance for policymakers, utilities, and homeowners. Our findings support informed decision-making and strategies tailored to the state's unique energy needs.
\end{itemize}

To address these questions, we conduct a comprehensive analysis of transforming Florida residences into NZEBs over the next 30 years. For existing homes, we evaluate the monthly cost savings of installing a 9.5 kW rooftop PV system, coupled with 42 kWh of battery storage to achieve average grid independence. We compare the financial benefits of owning rooftop PV systems with zero, 50\%, and 100\% effective energy storage—both with and without the 30\% Federal Income Tax Credit (ITC)—against the cost of purchasing electricity from the grid from 2020 to 2050. For new home constructions in Orlando, we assess the monthly savings from integrating energy efficiency improvements, PV installations, and battery storage, contrasting these with the costs of traditional grid-dependent homes. Additionally, we explore the potential of using an average 68.7 kWh electric vehicle (EV) as emergency backup power through V2H technology, examining its impact on monthly savings and energy reliability. Finally, we discuss the technical and economic advantages of NZEBs in significantly increasing residential renewable energy penetration and effectively eliminating residential load on the grid. These developments present substantial opportunities for both residents and utilities, enabling Florida to achieve net-zero emissions at minimal cost. Residents benefit from reduced monthly expenses, participation in future peer-to-peer energy exchanges, and access to reliable backup power. Utilities gain by reducing transmission losses, operating residential PV and battery systems as virtual power plants, and optimizing utility-scale solar for rapid EV charging and hydrogen (H\textsubscript{2}) production, thereby decarbonizing both the power and transportation sectors.

\section{Florida Residents' Electricity Consumption}\label{FLA_ElecConcumption_Existig}

\begin{figure}
\centering
\footnotesize
\captionsetup{singlelinecheck=false,font={footnotesize}}
	\includegraphics[width=3.5in]{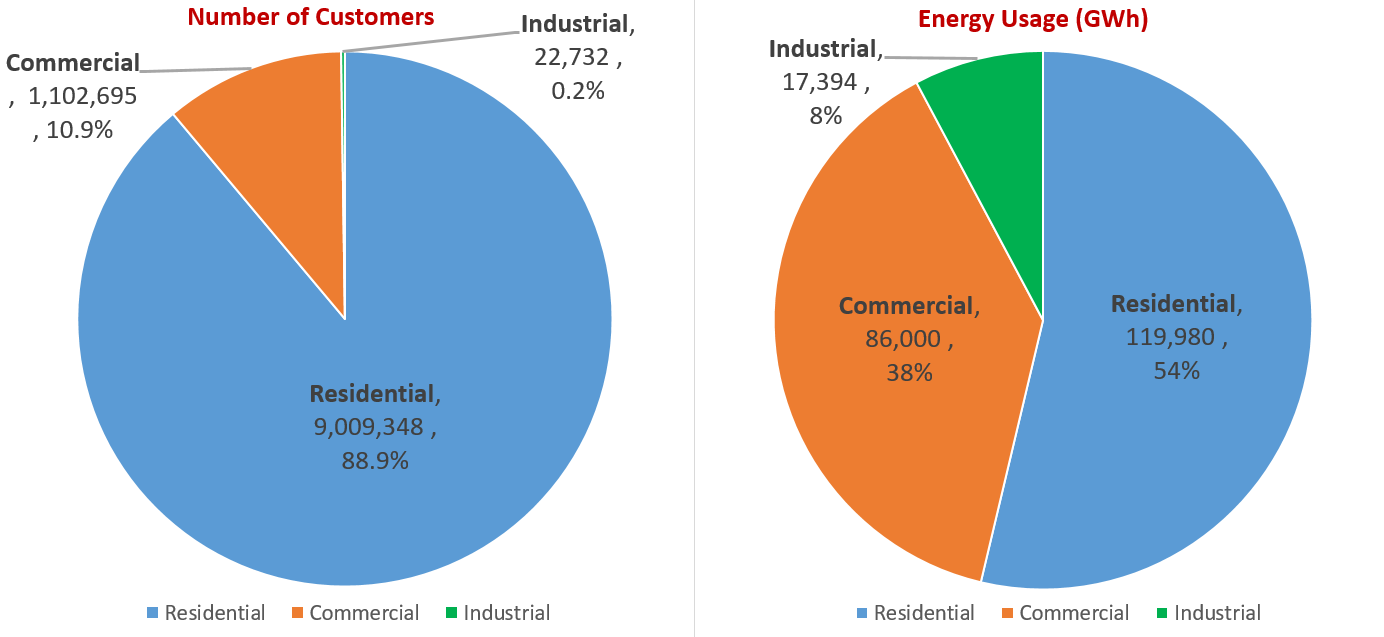}
	\caption{Florida Electric Customer Composition 2020}
    \label{FL_Electricity_Piecharts}
\end{figure}

Figure \ref{FL_Electricity_Piecharts}, based on data from the Florida Public Service Commission \cite{FLAPSC}, illustrates the composition of Florida's electric customer base in 2020. The left pie chart shows that residential customers constitute 89\% of the total customer base, emphasizing the dominance of the residential sector in the state's electricity market. The right pie chart indicates that residences consume 54\% of the state's electricity, while commercial establishments account for 39\%, demonstrating that over 90\% of Florida's electricity consumption occurs within buildings. To calculate the average electricity consumption per customer, we divide the total energy usage by the number of customers for each sector. For residential customers, this results in an average consumption of 13,300 kWh per year (approximately 1,100 kWh per month or 36 kWh per day). At an electricity rate of \$0.113 per kilowatt-hour, this translates to an average monthly bill of about \$124. Commercial customers consume an average of 77,000 kWh per year, whereas industrial customers use approximately 760,000 kWh per year, reflecting the higher energy demands of these sectors. Given these consumption patterns, if each residential customer were to install a 9.5 kW photovoltaic (PV) system—assuming an annual generation of 1,400 kWh per kW of PV capacity—they would produce approximately 13,300 kWh annually. This amount matches the average residential consumption, effectively transforming each home into a net-zero energy building. Such widespread adoption of residential PV systems could significantly reduce demand on the grid, lower greenhouse gas emissions, and contribute to Florida's renewable energy goals. This analysis underscores the substantial impact that residential energy efficiency and renewable energy integration can have on the state's overall electricity consumption and sustainability efforts.

\section{Simulation Inputs and Assumptions}
To conduct the techno-economic analysis and incorporate essential information, we used the parameters listed in Table \ref{parameters}. First, we focus on the utility energy delivery costs—including generation, transmission, distribution, taxes, and profits—and compare the rates for existing and new homes, both with and without PV systems and battery storage. In the second subsection, we present results from three case studies that demonstrate sufficient incentives (monthly savings on electricity demand and transportation) for customers to install PV systems, battery storage, and efficiency improvements, and to utilize their EVs not only to save money but also to have backup power during emergencies. 

\begin{table}[h]
\centering
\small 
\centering
\captionsetup{labelsep=space,font={footnotesize,sc}}
\caption{ \\ Key Parameters Used in Techno-Economic Analysis}
\label{parameters}
\centering
\begin{tabular}{|c|c|}
\hline
\vspace{0.2mm}
Existing Home PV (kW)         & 9.5    \\ \hline \vspace{0.2mm}
Existing Home PV with Improvement (kW)    & 6.48    \\ \hline\vspace{0.2mm}
PV Inverter Cost (\$/pWdc)            & 0.1    \\ \hline\vspace{0.2mm}
PV Inverter Life (Years)              & 15     \\ \hline\vspace{0.2mm}
PV Degradation (\%/year)              & 0.5    \\ \hline\vspace{0.2mm}
100\% Battery Capacity (kWh)          & 42.21  \\ \hline\vspace{0.2mm}
Battery Capacity for Improved Home (kWh) & 28.82  \\ \hline\vspace{0.2mm}
Battery Efficiency (\%)               & 95     \\ \hline\vspace{0.2mm}
Battery Degradation (\%)              & 3.5    \\ \hline\vspace{0.2mm}
Battery Life (Years)                  & 10     \\ \hline\vspace{0.2mm}
Grid Electricity Price (\$/kWh)       & 0.113  \\ \hline\vspace{0.2mm}
Analysis Period (Years)               & 30     \\ \hline\vspace{0.2mm}
Service Time (Years)                  & 25     \\ \hline\vspace{0.2mm}
Down Payment (\%)                     & 10     \\ \hline\vspace{0.2mm}
General Inflation Rate (\%)           & 2.5    \\ \hline\vspace{0.2mm}
Real Discount Rate (\%)               & 1.95   \\ \hline\vspace{0.2mm}
Nominal Discount Rate (\%)            & 4.5    \\ \hline\vspace{0.2mm}
Average EV Battery Size (kWh)                 & 68.7   \\ \hline\vspace{0.2mm}
Average EV Range (mile)                 & 220    \\ \hline\vspace{0.2mm}
Efficiency Improvement in Florida (\%)           & 31.7   \\ \hline\vspace{0.2mm}
Marginal Tax Rate (\%)                & 20     \\ \hline\vspace{0.2mm}
Real Electricity Price Escalation (\%)& 0      \\ \hline\vspace{0.2mm}
Nominal Electricity Price Escalation (\%) & 2.5 \\ \hline
\end{tabular}
\end{table}

\subsection{Utility Energy Delivery Cost Decomposition}
\begin{figure}
\centering
\footnotesize
\captionsetup{singlelinecheck=false,font={footnotesize}}
	\includegraphics[height=1.9in, width=3.5in]{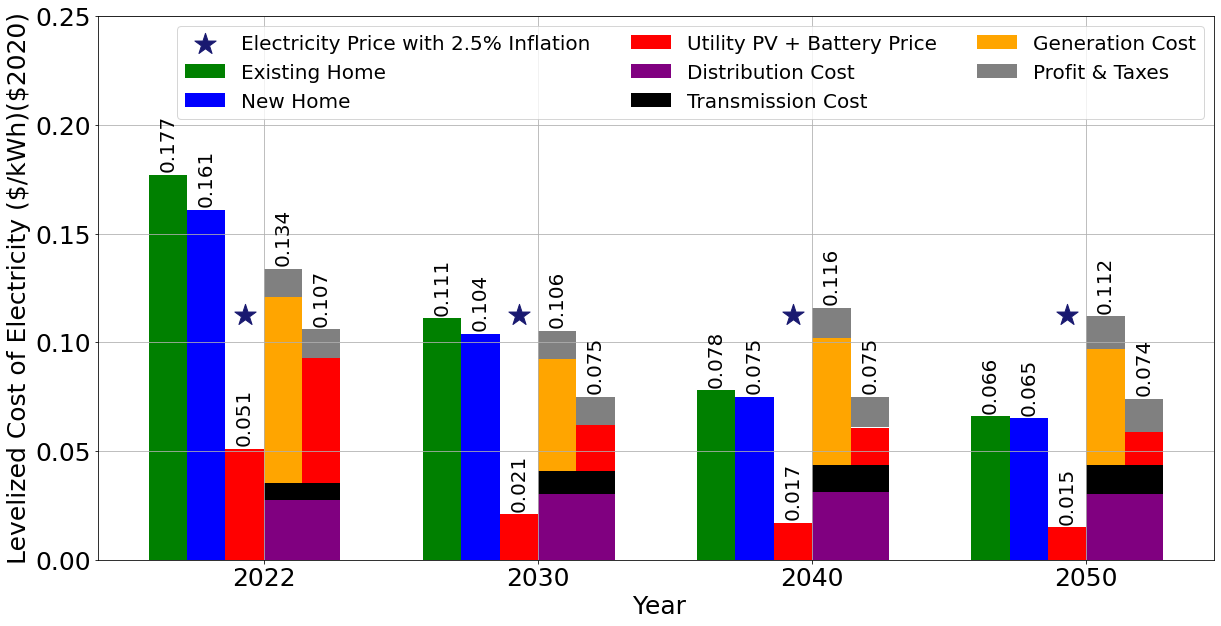}
	\caption{Florida Residential PV + Battery (100\% effective storage) and Utility PV + Battery (4-hour) "out of the wall" Cost of Electricity, without the ITC for PV and Battery}
    \label{LCOE_PV_Batt_Comparison}
\end{figure}

\begin{figure}
\centering
\footnotesize
\captionsetup{singlelinecheck=false,font={footnotesize}}
	\includegraphics[height=1.9in, width=3.5in]{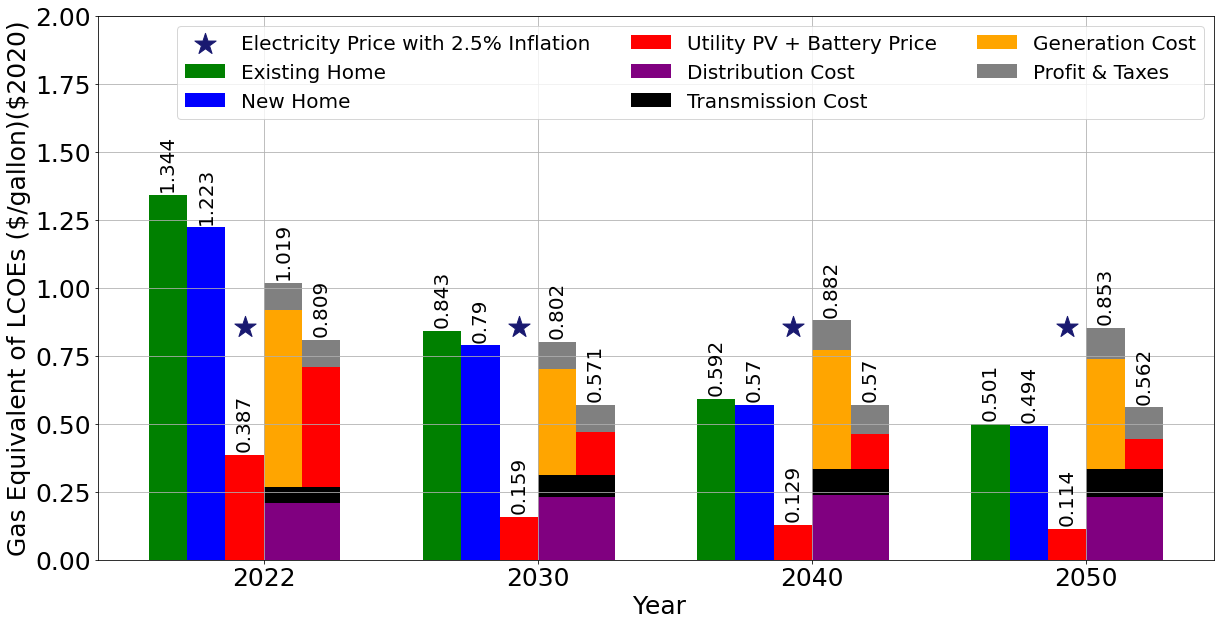}
	\caption{Gas Equivalent of LCOEs (\$/gallon) for PV+Battery Scenario without 30\% PV and Battery ITC.}
    \label{gas_equi_PVBatt}
\end{figure}

Figure \ref{LCOE_PV_Batt_Comparison} illustrates the levelized cost of electricity (LCOE) for existing homes and newly built homes equipped with PV systems and 4-hour lithium-ion batteries over the years 2022 to 2050, without applying investment tax credits (ITCs) on the PV and battery systems. Existing homes are equipped with a rooftop 9.5 kW PV system and a 42.2 kWh battery, while newly built homes have a rooftop 8.6 kW PV system and a 38.3 kWh battery. The actual utility electricity cost decomposition—including generation, transmission, and distribution costs—is consistently more expensive than the LCOE of PV and battery systems, even when considering transmission and distribution costs. This figure highlights that it is more cost-effective for utilities in Florida to install solar fields and equip them with battery storage in order to increase renewable penetration using these systems to generate green energy and deliver it to customers, thereby achieving net-zero emission policy targets by 2050. On the other hand, if customers equip their homes with PV and battery systems with mentioned capacities, they can become net-zero homes, save money by reducing reliance on expensive grid electricity, and contribute to environmental sustainability by reducing carbon emissions through their own green energy production. 
From a transportation perspective, Figure \ref{gas_equi_PVBatt} calculates the gasoline equivalent of these LCOE prices. It assumes that the average EV has a range of 220 miles and a 68.7 kWh battery, resulting in an efficiency of 3.20 miles per kWh. The average gasoline vehicle achieves 24.2 miles per gallon (mpg) \cite{USDOEEERE}. Therefore, the gasoline equivalent in dollars per gallon can be calculated by multiplying the price of electricity in dollars per kWh by the ratio of 24.2 mpg to 3.20 miles per kWh. The LCOE for rooftop solar in an existing home in 2022 (green column of Figure \ref{LCOE_PV_Batt_Comparison}) is 17.7 cents per kWh. This is equivalent to \$1.34 per gallon of gasoline, calculated as \$0.177 × 24.2 ÷ 3.20, where 24.2 mpg is the average efficiency of a gasoline vehicle \cite{USDOEEERE}, and 3.20 miles per kWh is the efficiency of an average EV with a 220-mile range and a 68.7 kWh battery. This calculation is presented in Figure \ref{gas_equi_PVBatt}. Considering that the average price for gasoline in 2023 is \$3.56 per gallon in 2023 dollars \cite{GasFLPrice}, or \$3.16 per gallon in 2020 dollars, the gasoline equivalent cost of rooftop PV is approximately 42\% of the cost of gasoline.

\subsection{Monthly Savings for FL Residents}
In this section, we examine the incentives for Florida residents to install photovoltaic (PV) systems and batteries, highlighting how much they can save by reducing reliance on the grid and becoming net-zero homes. Installing a battery not only contributes to energy savings but also provides backup power during emergencies. We present different case studies that detail the monthly savings for existing homes in Florida under various scenarios: installing only PV, adding a battery, implementing efficiency improvements, and incorporating an electric vehicle (EV).

\subsubsection{Monthly Savings with PV and Battery}
\begin{figure}
\centering
\footnotesize
\captionsetup{singlelinecheck=false,font={footnotesize}}
	\includegraphics[height=1.9in, width=3.5in]{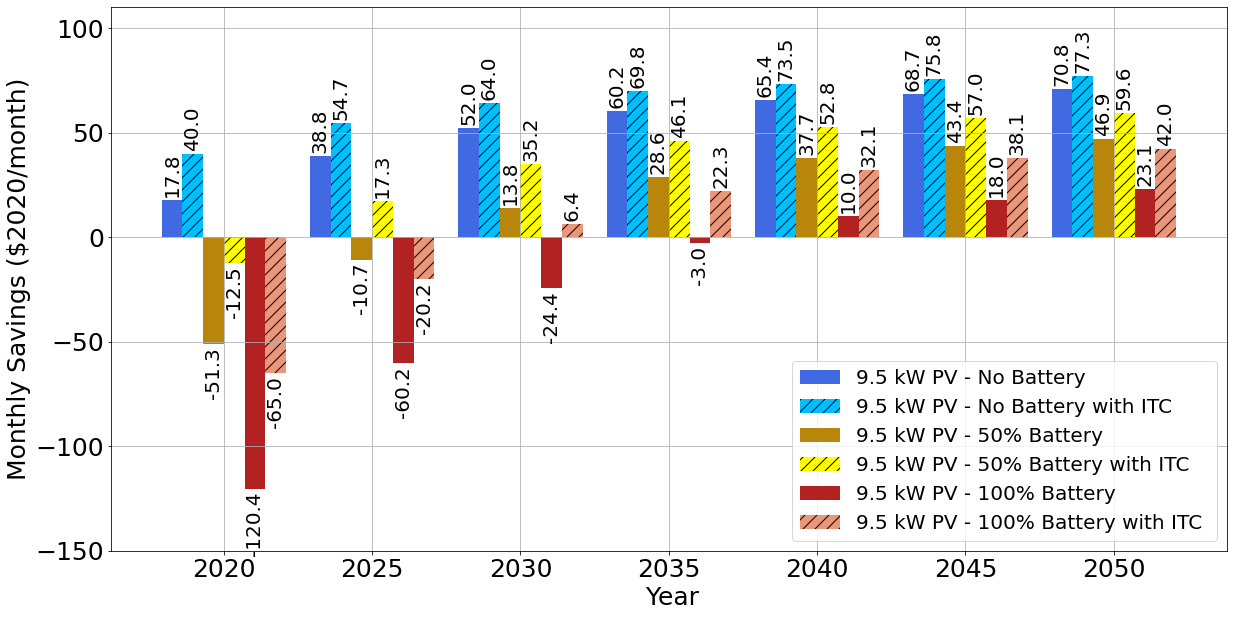}
	\caption{Monthly Savings with PV and Batteries Making the Average Florida Residence Net Zero}
    \label{Monthly_Saving_Battery_ITC}
\end{figure}

Figure \ref{Monthly_Saving_Battery_ITC} illustrates the monthly savings for an existing Florida resident equipped with a 9.5 kW photovoltaic (PV) system and various battery capacities (with 100\% reference being 42.2 kWh), both with and without Investment Tax Credits (ITCs). Without a battery, the resident can save approximately \$18 per month without ITCs, increasing to \$40 per month with ITCs. However, when batteries are included, the monthly savings become negative in the first years, indicating a financial loss primarily due to the upfront installation costs of the batteries. Nevertheless, as shown, residents can achieve grid independence by 2037 without ITCs and by 2029 with ITCs, with having both PV and battery respectively. The key point here is that incorporating batteries not only enables residents to make their homes grid-independent but also enhances resilience against power outages resulting from hurricanes and floods in the area. 

\subsubsection{Monthly Savings with PV, Battery, EV with Efficiency Improvement}
\begin{figure}
\centering
\footnotesize
\captionsetup{singlelinecheck=false,font={footnotesize}}
\includegraphics[height=1.9in, width=3.5in]{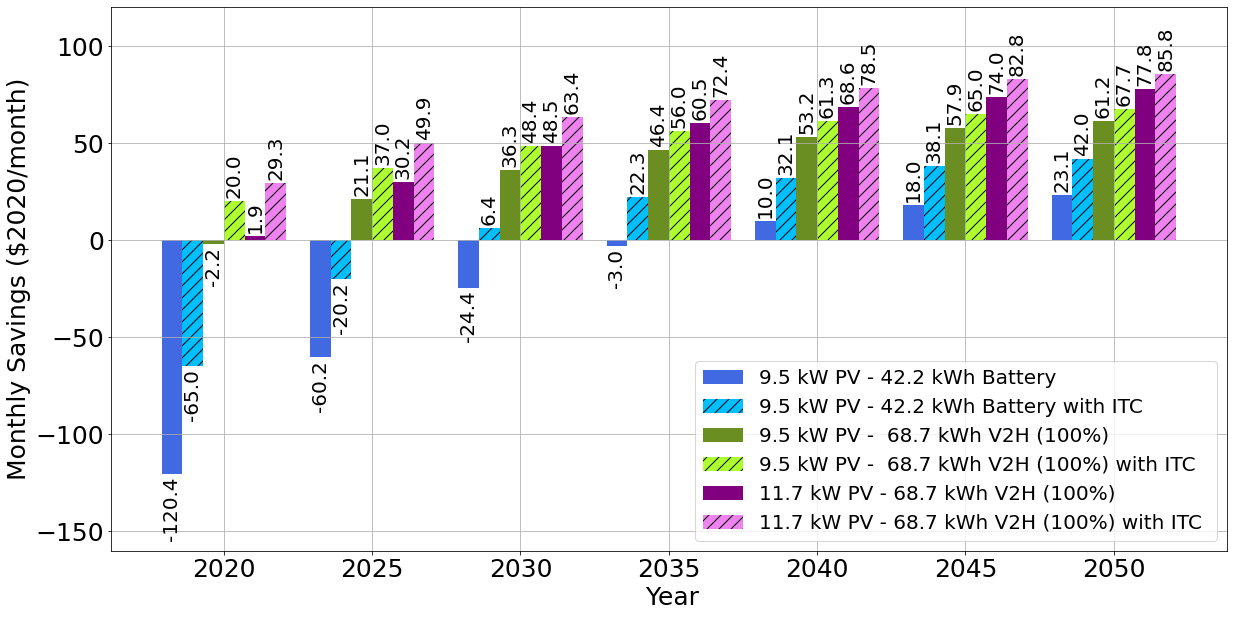}
\caption{Monthly Savings with PV, Batteries, and V2H with and without ITC Making the Average Florida Residence Net Zero}
\label{V2G_IncomeTax}
\end{figure}

\begin{figure}
\centering
\footnotesize
\captionsetup{singlelinecheck=false,font={footnotesize}}
	\includegraphics[height=1.9in, width=3.5in]{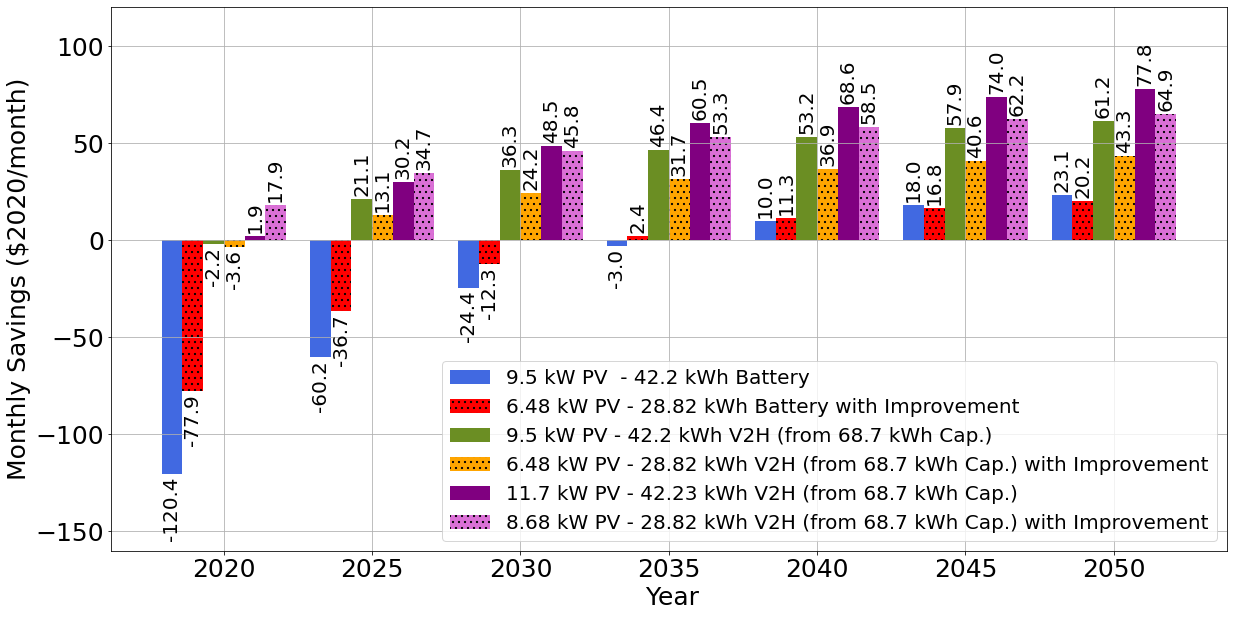}
	\caption{Monthly Savings with PV, Batteries, and V2H with Efficiency Improvement Making the Average Florida Residence Net Zero}
    \label{V2G_improvement}
\end{figure}

To assess the impact of V2H capability and also energy efficiency improvement of existing homes and how Florida residents reduce electricity bills and save money, Figure \ref{V2G_IncomeTax} and Figure \ref{V2G_improvement} are shown.  Considering the relatively low upfront cost of the bidirectional charger (\$6,000) compared to wall batteries, all of the homes in Figure \ref{V2G_IncomeTax} can cost-effectively provide backup power in 2021 using existing PV on the roof and the homeowner’s EV  well before the bi-directional chargers and EVs are ready for V2H in 2024. The addition of 2.2 kW of PV to the 9.5 kW home with V2H capability provides extra electricity, enabling an additional 3,080 kWh/yr or approximately 10,000 miles/year of driving powered by solar electricity. 
Figure \ref{V2G_improvement} also shows how home efficiency improvements can play an important role in monthly savings. With efficient homes there is no need to add 9.5 kW PV to make the home net zero. Instead, with lower PV size such as 6.48 kW, residents can still win on paying less on investment of PV (and accordingly lower battery size) and save almost the same as previous scenarios with battery. For instance, 8.68 kW PV with lower battery size from EV and with efficiency improvement saves more money compared to 11.7 kW PV with an even larger battery size.

\subsubsection{Monthly Savings with PV, Battery, EV, with PV+Battery ITCs, with Gasoline Savings}

\begin{figure}
\centering
\footnotesize
\captionsetup{singlelinecheck=false,font={footnotesize}}
	\includegraphics[height=1.9in, width=3.5in]{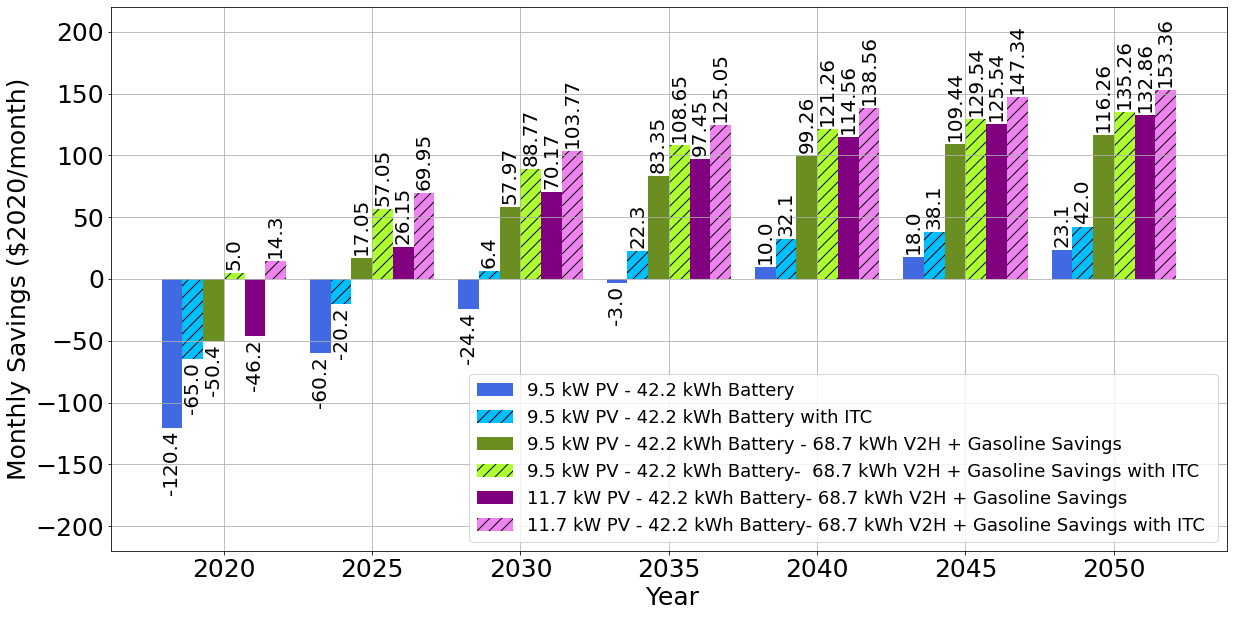}
	\caption{Monthly Savings with adding PV, Batteries, and V2H Capability and Gasoline Savings with and without ITC Making the Average Florida Residence Net Zero}
    \label{V2H_transportationsaving_extraPV}
\end{figure} 

Figure \ref{V2G_IncomeTax} presents the projected monthly savings in 2020 USD for different photovoltaic (PV) and battery system configurations from 2020 to 2050. The analysis includes multiple PV setups, such as 9.5 kW PV with a 42.2 kWh battery, 9.5 kW PV with vehicle-to-home (V2H) integration, and 11.7 kW PV with V2H. Additionally, the impact of the Investment Tax Credit (ITC) is evaluated for these configurations. In Figure \ref{V2G_IncomeTax}, two primary setups are presented. The first involves 9.5 kW of PV with 100\% effective storage using a 42.2 kWh battery (dark and light blue columns). The second scenario features 9.5 kW of PV with a bidirectional charger (\$6,000 installed cost) and a V2H-capable electric car (dark and light green columns). Additionally, there is a setup with 11.7 kW of PV, a bidirectional charger, and a V2H-capable electric car (dark and light purple columns).

Initially, in 2020, most systems present negative monthly savings, indicating higher costs compared to the baseline without PV or storage. The 9.5 kW PV system with a 42.2 kWh battery, without ITC, has the highest losses, with monthly savings of -\$120.4. However, the ITC significantly mitigates these losses, reducing them to -\$65.0 per month. The upfront cost of wall-mounted batteries for 100\% storage is cost-effective for backup power, especially before 2030, due to the ITC. In contrast, the relatively low cost of the bidirectional charger (\$6,000) compared to wall batteries makes V2H systems cost-effective as early as 2021, utilizing existing PV and an EV. This is feasible even before bidirectional chargers and EVss become widely available for V2H functionality in 2024. As time progresses, technological advancements lead to reductions in PV and battery costs, as well as improved efficiencies and decreased battery degradation rates. By 2030, most configurations yield positive savings, with the 9.5 kW PV system with 68.7 kWh V2H (including ITC) achieving the highest monthly savings of \$48.5. V2H technology provides flexibility by allowing stored energy to be used during peak demand periods, when grid electricity is most expensive, thereby improving overall savings.

The larger PV system with 11.7 kW capacity consistently outperforms the smaller systems. By 2050, the 11.7 kW PV system with 68.7 kWh V2H and ITC will achieve the highest savings of \$85.8 per month. The larger PV capacity generates more electricity, which can be used to meet household demand or stored for later use, further reducing reliance on grid power and increasing the potential for surplus energy export under feed-in tariffs or net metering. The inclusion of V2H capability further enhances economic viability by offering significant additional savings beyond energy cost reductions. For example, adding 2.2 kW of PV to a 9.5 kW system with V2H generates an additional 3,080 kWh per year, enough for approximately 10,000 miles of driving on solar electricity. The cost of driving 10,000 miles on solar power is estimated at \$124.92 per year, compared to \$1,305.80 per year for gasoline, resulting in annual savings of \$1,180.90, or \$98.40 per month. The presence of ITC consistently improves economic feasibility across all configurations, highlighting the importance of financial incentives in promoting renewable energy adoption. 

\begin{figure}
\centering
\footnotesize
\captionsetup{singlelinecheck=false,font={footnotesize}}
\includegraphics[height=1.9in, width=3.5in]{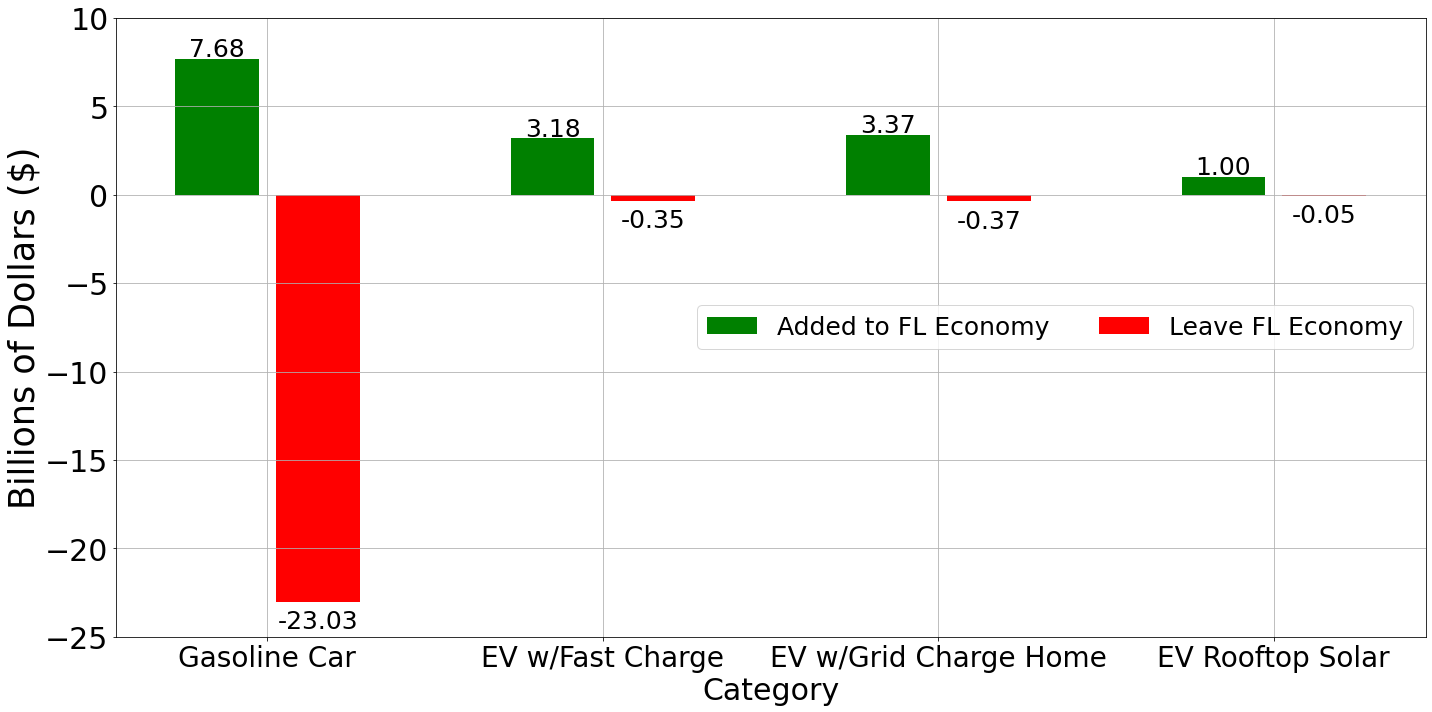}
\caption{Florida economics for gasoline car vs EV car with different charging options.}
\label{FloridaSavemoney}
\end{figure}

From both a transportation standpoint and a broader statewide perspective, Figure \ref{FloridaSavemoney} shows the large economic loss and energy waste from using gasoline in Florida. In 2022, Florida spent \$30.7 billion on gasoline, and 75\% of that (\$23.03 billion) left the state because all gasoline was imported. Gasoline engines are also very inefficient, with only 21\% of the energy used for driving and the rest wasted. If gasoline vehicles were replaced by EVs, approximately \$23 billion could be retained in the state of Florida and used for other important needs, boosting the state’s economy and reducing waste.

\section{Conclusion}\label{conclusion}
This paper conducted a 30-year techno-economic analysis of Net-Zero Energy Buildings (NZEBs) in Florida, focusing on both existing and new homes in Orlando. The financial benefits of combining rooftop solar panels, battery storage, and electric vehicles (EVs) were evaluated in comparison to relying on grid electricity. By utilizing federal incentives like the Investment Tax Credit and implementing energy efficiency improvements, residents could achieve significant cost savings. Simulations based on real energy consumption data from Florida homes showed that existing homes with a 9.5 kW solar system and a 42.2 kWh battery could begin generating positive returns by 2029, while new homes could reach this point as early as 2024. Additionally, charging an EV with rooftop solar energy could save residents up to \$100 per month compared to purchasing fuel for an average gasoline car. Integrating solar panels and battery storage not only reduced electric bills but also increased grid independence and resilience during power outages. On a broader scale, NZEBs contributed to grid stability by decreasing electricity demand and supporting the adoption of utility-scale renewable energy. This approach could lower infrastructure costs, help Florida align with national decarbonization goals, retain approximately \$23 billion within the state, and advance progress toward a sustainable, low-carbon future.



\setstretch{0.99}

\bibliographystyle{IEEEtran}
\bibliography{mybib}

\end{document}